\documentclass[English,superscriptaddress,prl,twocolumn]{revtex4-1}
\usepackage{graphicx} 
\usepackage[colorlinks=true,linkcolor=blue,urlcolor=blue,citecolor=blue]{hyperref}
\usepackage{xcolor}
\usepackage{soul}
\usepackage[normalem]{ulem}

\begin{document}

\title{Twist-Angle-Controlled Anomalous Gating in Bilayer Graphene/BN Heterostructures}

\author{G. Maffione}
\affiliation{Universit\'e Paris-Saclay, CNRS, Centre de Nanosciences et de Nanotechnologies (C2N), 91120 Palaiseau, France}

\author{L. S. Farrar}
\affiliation{Universit\'e Paris-Saclay, CNRS, Centre de Nanosciences et de Nanotechnologies (C2N), 91120 Palaiseau, France}

\author{M. Kapfer}
\affiliation{Universit\'e Paris-Saclay, CNRS, Centre de Nanosciences et de Nanotechnologies (C2N), 91120 Palaiseau, France}

\author{K. Watanabe}
\affiliation{Research Center for Electronic and Optical Materials, National Institute for Materials Science, 1-1 Namiki, Tsukuba 305-0044, Japan
}
\author{T. Taniguchi}
\affiliation{Research Center for Materials Nanoarchitectonics, National Institute for Materials Science,  1-1 Namiki, Tsukuba 305-0044, Japan}

\author{H. Aubin}
\affiliation{Universit\'e Paris-Saclay, CNRS, Centre de Nanosciences et de Nanotechnologies (C2N), 91120 Palaiseau, France}

\author{D. Mailly}
\affiliation{Universit\'e Paris-Saclay, CNRS, Centre de Nanosciences et de Nanotechnologies (C2N), 91120 Palaiseau, France}
\author{R. Ribeiro-Palau*}
\affiliation{Universit\'e Paris-Saclay, CNRS, Centre de Nanosciences et de Nanotechnologies (C2N), 91120 Palaiseau, France}

\newcommand{\missing}{\textcolor{orange}}

\newcommand{\Dom}{\textcolor{blue}}
\newcommand{\Gaia}{\textcolor{black}}

\newcommand{\Reb}{\textcolor{purple}}

\maketitle

\noindent \textbf{Anomalous gating effects—such as gate ineffectiveness and pronounced hysteresis—have been observed in graphene-based systems encapsulated in boron nitride (BN) and linked to a possible ferroelectric state. However, their origin, stability, and reproducibility remain under debate. Here, we present charge transport experiments in dual-gated, dynamically rotatable van der Waals heterostructures based on bilayer graphene encapsulated in BN. Remarkably, the angular degree of freedom acts as an ON/OFF switch for the anomalous gating response. We show that the angular alignment between the two BN layers—not the presence of a moir\'e superlattice with graphene—is the key parameter governing these effects.  The relevant alignment between the two BN layers, to observe the anomalous gating effect at room temperature, lies between 15$^{\circ}$ and 45$^{\circ}$, with no evidence of the expected 60$^{\circ}$ periodicity. Both gate ineffectiveness and hysteresis are highly sensitive to small angular changes, which we classify into three distinct regimes. Our results clarify the conditions necessary to reproduce these phenomena and pave the way for theoretical investigation of their microscopic origins.}


Crystallographic alignment in van der Waals heterostructures has been proven to be an utmost important parameter, whose consequences need to be thoroughly understood before considering their use in future applications. This is also the case for the recently observed ineffectiveness and hysteretic behavior of the electrostatic gates in graphene/BN structures, associated with a ferroelectric effect \cite{lin_room_2025,zhang_electronic_2024,zheng_unconventional_2020,niu_giant_2022,singh_stacking-induced_2025,kleinElectricalSwitchingBistable2023,chenAnomalousGatetunableCapacitance2024,waters_anomalous_2025,niu_ferroelectricity_2025}. Such a ferroelectric effect could lead to the development  of ferroelectric field effect transistors, where the electrical conductance and threshold voltage are controlled by the switching of the electrical polarization of the ferroelectric layer\cite{yasudaUltrafastHighenduranceMemory2024,yan2023moire}.  However, although gate ineffectiveness  and hysteretic behavior have now been observed in encapsulated graphene systems across various structures — including monolayer \cite{lin_room_2025,zhang_electronic_2024}, bilayer \cite{zheng_unconventional_2020,niu_giant_2022,zhang_electronic_2024}, trilayer \cite{zhang_electronic_2024}, tetralayer \cite{singh_stacking-induced_2025}, magic-angle twisted bilayer \cite{kleinElectricalSwitchingBistable2023}, double twisted bilayer \cite{chenAnomalousGatetunableCapacitance2024}, twisted multilayer \cite{waters_anomalous_2025}, and graphene semi-encapsulated on WSe$_{2}$ and BN \cite{niu_ferroelectricity_2025} — a real understanding of the observed phenomena is missing due to a lack of control of the sample alignment.

Here, we present charge transport measurements performed in dynamically rotatable, dual-gated van der Waals heterostructures, which allow us to activate and deactivate the anomalous gating behavior. We demonstrate that in order to turn ON the anomalous gating, at room temperature, it is necessary to have a rotational alignment between top and bottom BN flakes between 15$^{\circ}$ and 45$^{\circ}$, irrespective of the angle with the graphene layer. We identify three types of behavior of the anomalous gating, which we separate according to their response to top and bottom gate voltage. 


\begin{figure*}[t]
    \centering
    \includegraphics[width = 1\textwidth]{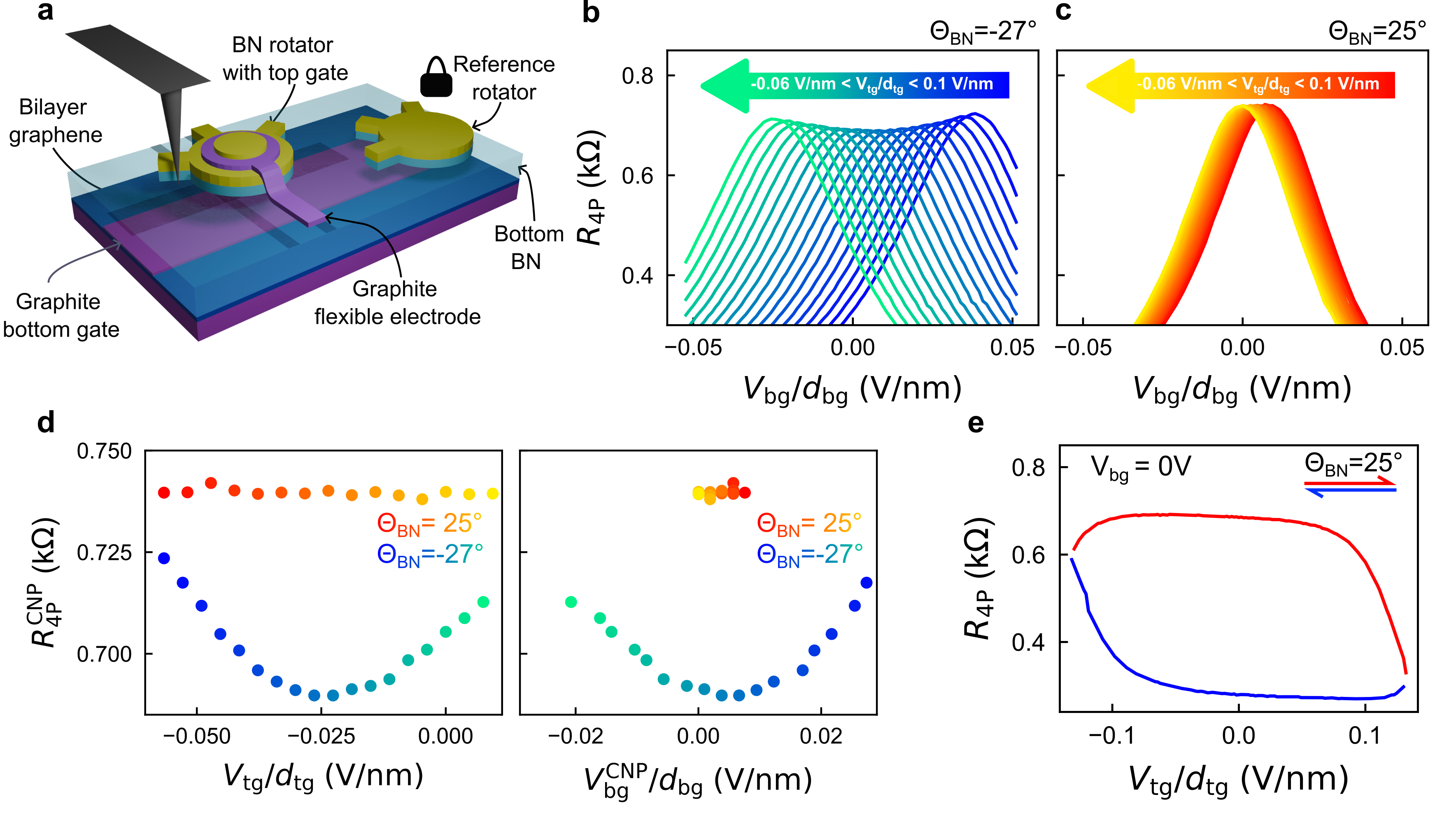}
    \caption{\textbf{Angle controlled anomalous gating.} \textbf{a}, Schematics of the dynamically rotatable van der Waals heterostructure with top gate and the reference rotator. \textbf{b} and \textbf{c}, four-probe resistance measurement, $R_{\mathrm{4P}}$, as a function of the bottom gate voltage, $V_{\mathrm{bg}}$, for several top gate values, $V_{\mathrm{tg}}$, at room temperature for angular alignments of $\Theta_{\mathrm{BN}}=-27^{\circ}$ and $\Theta_{\mathrm{BN}}=25^{\circ}$, respectively. The gate voltages  $V_{\mathrm{bg}}$ and $V_{\mathrm{tg}}$ are normalized by the respective BN thickness $d_{\mathrm{bg}}$ and $d_{\mathrm{tg}}$. Top gates values for b and c go from $V_{\mathrm{tg}}= 0.01$ V/nm to $V_{\mathrm{tg}}= -0.06$ V/nm  in steps of $0.004$ V/nm. Bottom gate voltage sweeping range is the same for both figures.  \textbf{d}, Position of the CNP extracted from figure b (blue) and c (orange) as a function of the applied top (left) and bottom (right) gate voltage. \textbf{e}, Four-probe resistance measurement as a function of the normalized top gate voltage for $V_{\mathrm{bg}}=0$ V, measured for $\Theta_{\mathrm{BN}}=25^{\circ}$. Red (blue) line indicated the forward (backward) measurement. All measurements are performed at room temperature.}
    \label{fig:Figure_1}
\end{figure*}

We have fabricated dual-gated rotatable devices\cite{Farrar2025Jan} made of Bernal stacked bilayer graphene (BBG) semi-encapsulated on BN.  A full description of the sample fabrication process is detailed in the supplementary note 1. In these structures we have a rotational degree of freedom from the top BN, which we denote as the active rotator,  which is able to apply a vertical electric field to the heterostructure, Fig. \ref{fig:Figure_1}a. This device architecture allows us to probe various angular configurations in the same heterostructure, within a range of $\approx 140^{\circ}$. Additionally, we have added a reference rotator, placed directly on the bottom BN flake,  made from the same BN flake as the active rotator, Fig. \ref{fig:Figure_1}a.  We rotated the reference rotator until it reached the locked position, where it no longer moves. This position corresponds to a crystallographic alignment between the two BN layers of $0^{\circ}/60^{\circ}$. By comparing the angles between the reference and the active rotators we can extract the angle between the two BN layers $\Theta_{\mathrm{BN}}$. Nonetheless, this technique does not allow us to distinguish parallel (AA - $\Theta_{\mathrm{BN}}=0^{\circ}$) from antiparallel (AA' - $\Theta_{\mathrm{BN}}=60^{\circ}$) alignment. We therefore use the convention that the reference rotator is locked at $\Theta_{\mathrm{BN}}=0^{\circ}$ and the angles reported here are measured with respect to this one. Positive angles reflect a clockwise rotation and negative angles an anticlockwise rotation.

Using the reference rotator we are also able to identify the alignment between the bottom BN and the graphene, labeled $\theta_{\mathrm{bBN-G}}$. Most of the measurements reported here were performed at room temperature on sample 1, for which  $\theta_{\mathrm{bBN-G}}\approx10^{\circ}$. Samples 2 ($\theta_{\mathrm{bBN-G}}\approx30^{\circ}$) and 3 ($\theta_{\mathrm{bBN-G}}\approx20^{\circ}$) are used for comparison and low temperature measurements. A detailed description of all samples can be found in the supplementary note 2.

\begin{figure*}[t]
    \centering
    \includegraphics[width = 1\textwidth]{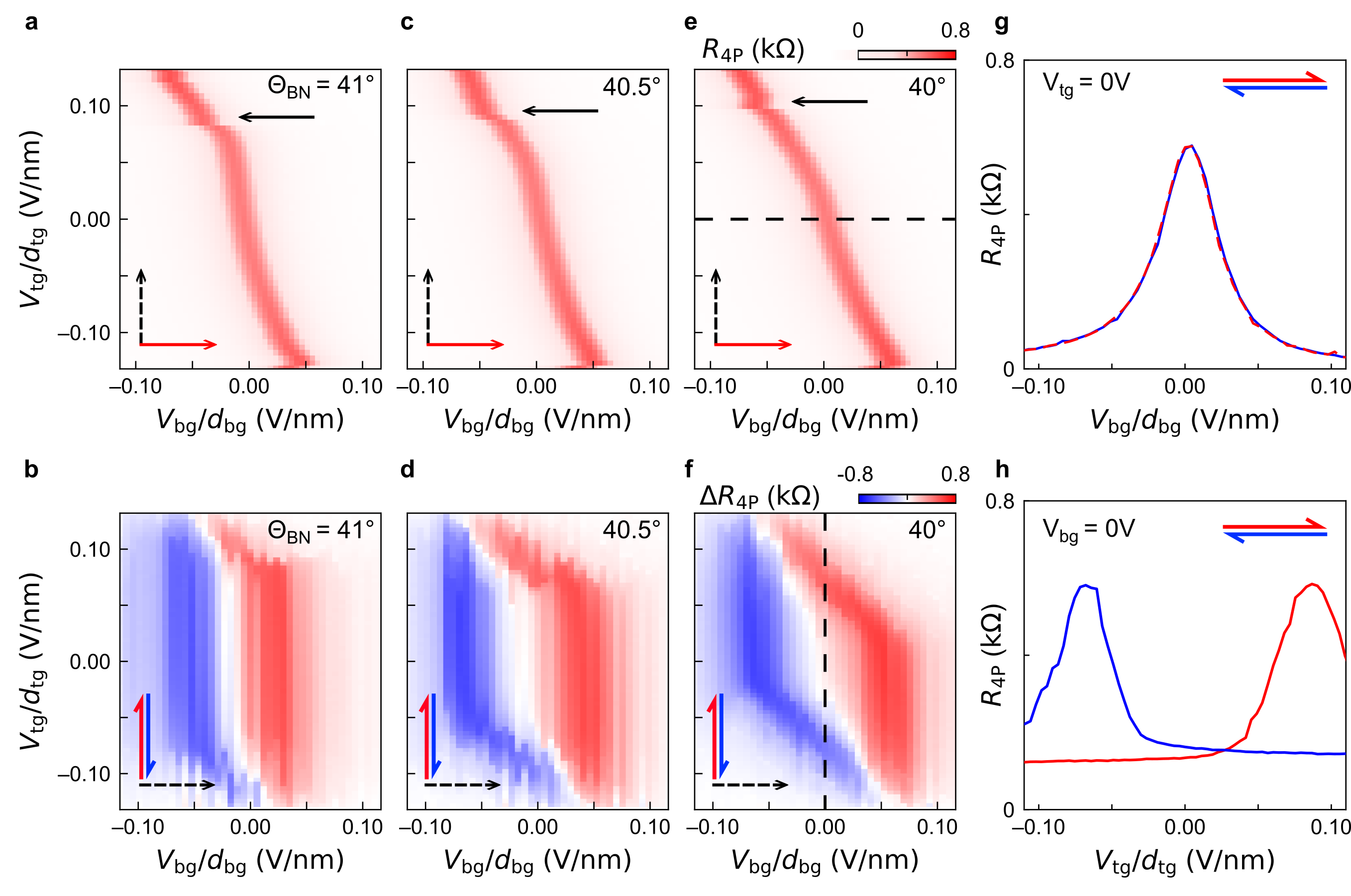}
    \caption{\textbf{Type I behavior.} \textbf{a, c} and \textbf{e}, Color-maps of the four-probe resistance as a function of the bottom gate voltage (fast axis) and the top gate voltage (slow axis), for three different angles, $\Theta_{\mathrm{BN}}=41^\circ,~ 40.5^\circ$ and $40^\circ$, respectively. The gate voltages are normalized by the thickness of their respective BN dielectrics. The three color-maps share the same color scale indicated on top of figure e.  \textbf{b, d} and \textbf{f}, Subtraction of four-probe resistance for the forward and backward traces of the top gate (fast axis) as a function of the bottom gate (slow axis) for the same three angular alignments. The three color-maps share the same color scale indicated on top of figure f. \textbf{g}, Four-probe resistance as a function of the normalized bottom gate voltage at $V_{\mathrm{tg}}=0$ V (dashed horizontal line in e). Red (dashed) and blue (solid) lines are the forward and backward traces, respectively. \textbf{h}, Four-probe resistance as a function of the normalized top gate voltage at $V_{\mathrm{bg}}=0$ V (vertical dashed line in f). Red and blue lines are the forward and backward traces, respectively. All measurements are  taken at room temperature. }
    \label{fig:Figure_2}
\end{figure*}

In Fig. \ref{fig:Figure_1}b we show  four-probe resistance plots as a function of the bottom gate voltage for $\Theta_{\mathrm{BN}}=-27^\circ$. In the following we present all the gate voltages normalized by the thickness of their respective BN dielectrics. Each curve is measured by sweeping the bottom gate voltage at a fixed top gate voltage from $-0.06$ V/nm to 0.01 V/nm in steps of 0.004 V/nm. We observe the standard behaviour of bilayer graphene, whereby varying the top gate voltage alters the resistance of the charge neutrality point (CNP) and causes a linear shift of its position with respect to the bottom gate voltage. The observed increase in the resistance is due to the induced potential difference between the layers of BBG, which opens an energy gap in the electronic band structure \cite{taychatanapat2010electronic, icking2022transport}. The linear shift of the position of the CNP is due to the change in carrier density induced by the top gate voltage which needs to be compensated by a proportional bottom gate voltage in order to reach zero carrier density. This standard behavior is observed for most accessible angular alignments, for more details see supplementary note 4.

As we rotate toward positive angles, signs of gate ineffectiveness begin to emerge at $\Theta_{\mathrm{BN}}=25^\circ$. Figure \ref{fig:Figure_1}c presents four-probe resistance plots measured at this angle in the same voltage range as for Fig. \ref{fig:Figure_1}b. We observe that the resistance at the CNP remains constant and the shift of the bottom gate voltage induced by the top-gate voltage is much smaller than in the standard gating behavior, reflecting the ineffectiveness of the top gate.

By extracting the resistance values at the CNP as a function of the applied top and bottom gate voltage for $\Theta_{\mathrm{BN}}=-27^\circ$ and $\Theta_{\mathrm{BN}}=25^\circ$, Fig. \ref{fig:Figure_1}d, the difference between the two angular alignments is clearly pointed out. Concerning the resistance of the CNP we observed the standard gating behavior described above for $\Theta_{\mathrm{BN}}=-27^\circ$. However, at $\Theta_{\mathrm{BN}}=25^\circ$ the resistance of the CNP is independent of the applied top gate voltage. This behavior is reminiscent of the one of monolayer graphene, where the applied displacement field does not open an energy gap and the use of two gates does not have an impact on the resistance of the CNP. We also note that for $\Theta_{\mathrm{BN}}=25^{\circ}$, the resistance value at which the CNP is pinned appears to be higher than the one with standard gating behavior. We expect these two angles to exhibit a similar resistance at the CNP given their similar crystal field value, a consequence of the atomic configuration of the top and bottom BNs \cite{Farrar2025Jan}. The difference in the resistance of the CNP could then be related to a strong intrinsic potential difference between the layers of BBG. We note that this does not seem to generate an intrinsic doping on the sample as the CNP is still observed close to zero bottom gate voltage. 

Plotting the resistance of the CNP as a function of its positions in bottom gate voltage (Fig. \ref{fig:Figure_1}d-right), shows that the voltage excursion of the bottom gate voltage is very small, indicating an ineffective top gate. This ineffectiveness of the top gate can also be observed when we sweep the top gate voltage, Fig. \ref{fig:Figure_1}e, for $V_{\mathrm{bg}}=0$ V. In this case, the value of the resistance is pinned at that of the CNP over a wide range of top gate voltage. By sweeping the top gate in both directions (forward and backward), we observe a strong hysteretic behavior. This hysteretic effect is present for any sweeping range of top gate voltage and is absent when we sweep the bottom gate (see supplementary note 5). The differences between Fig. \ref{fig:Figure_1}c and e highlight a significant gate asymmetry and underscore the importance of independently exploring both gates.

\begin{figure*}
    \centering
    \includegraphics[width = 1\textwidth]{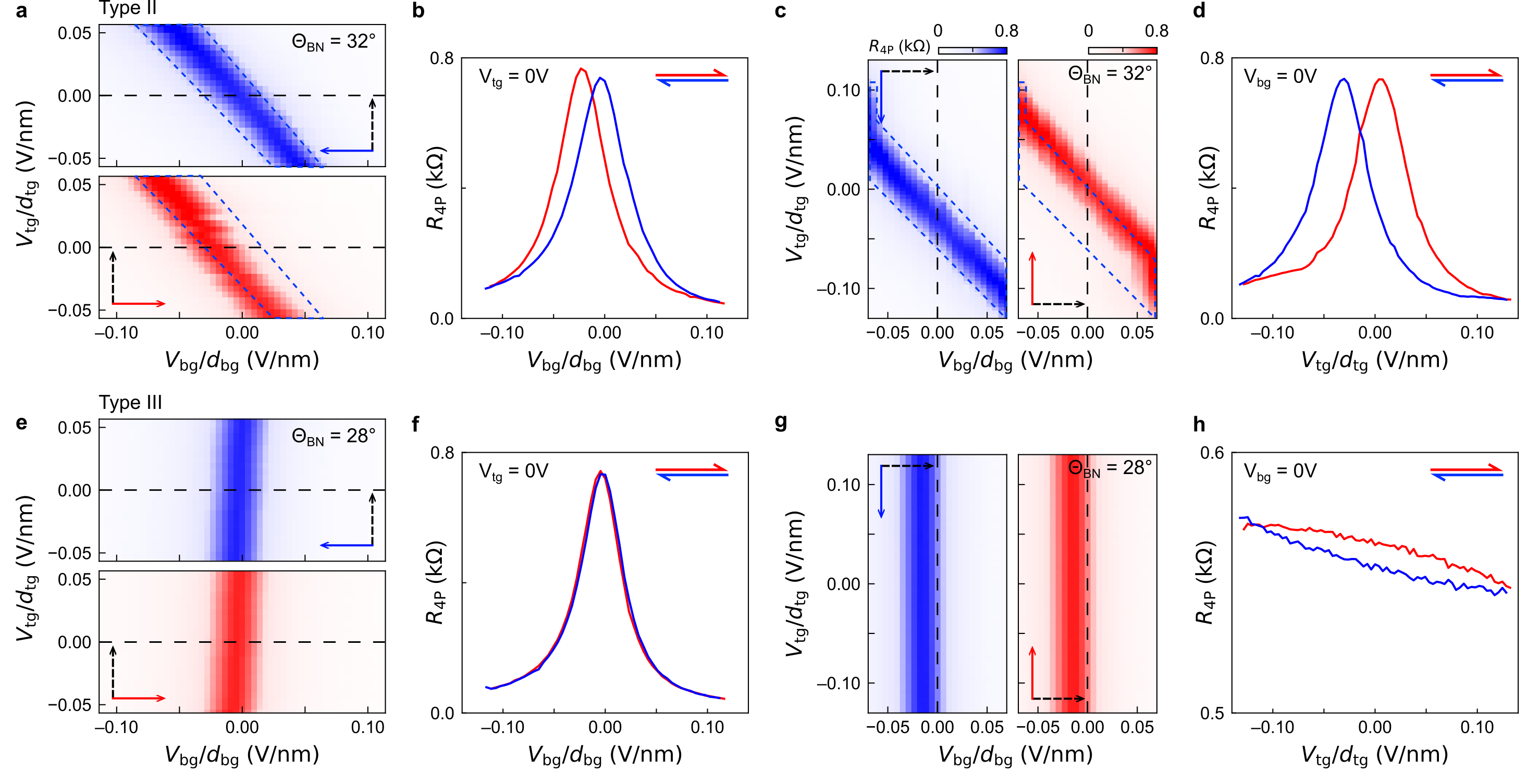}
    \caption{\textbf{Type II and III behaviors.} \textbf{a}, Color-maps of the four-probe resistance as a function
of the forward (bottom) and backward (top) traces of the normalized bottom (fast axis) and top gate voltage (slow axis) for $\Theta_{\mathrm{BN}}=32^{\circ}$. Dashed blue line indicates the position of the CNP for the backward direction. \textbf{b}, Four-probe resistance as a function of the normalized bottom gate voltage at $V_{\mathrm{tg}}=0$ V (horizontal black dashed lines in a). Red and blue  lines are the forward and backward traces, respectively. \textbf{c}, Color-maps of the four probe resistance as a function
of the forward (right) and backward (left) traces of the normalized top (fast axis) and bottom gate voltage (slow axis)  for the same angular alignment. Dashed blue line indicates the position of the CNP for the backward direction. \textbf{d}, Four-probe resistance as a function of the normalized top gate voltage at $V_{\mathrm{bg}}=0$ V (vertical black dashed lines in a). Red and blue  lines are the forward and backward traces, respectively. \textbf{e, f, g} and \textbf{h}, Same as a, b, c and d, respectively,  for $\Theta_{\mathrm{BN}}=28^{\circ}$.  All measurements are taken at room temperature. All color-maps share the same color scale indicated on top of figure c.}
    \label{fig:Figure_3}
\end{figure*}

To better understand the anomalous gating behavior we study the electronic transport characteristics through color-maps of the four-probe resistance for various angles around $\Theta_{\mathrm{BN}}=30^\circ$. Each 2D-map is recorded as follows: a curve is traced by sweeping one gate back and forth (fast axis, marked by a solid arrow pointing in the direction of the sweep) and once this is completed the value of the other gate is  increased step-like (slow axis, marked by dashed arrows). We repeat this procedure until the full 2D-map is completed. In this report we always keep the same orientation of the color-maps, whereby the bottom gate is on the $x$-axis and top gate is on the $y$-axis. The direction of slow (dashed) and fast (solid) axis is clearly identified in each map by arrows located close to the corner where the map starts. By comparing the color-maps of the resistance at different angular positions we distinguish three predominant behaviors, which we will refer to as type I, type II and type III. 


\noindent \textbf{Anomalous gating type I:} Figure \ref{fig:Figure_2}a, c and e  shows  the four-probe resistance as a function of  $V_{\mathrm{bg}}/d_{\mathrm{bg}}$ (forward trace - fast axis) and $V_{\mathrm{tg}}/d_{\mathrm{tg}}$ (slow axis) for three different angular alignments $\Theta_{\mathrm{BN}}=41^{\circ}$, 40.5$^{\circ}$ and 40$^{\circ}$ (where the measurements were recorded consecutively). Figure \ref{fig:Figure_2}g, shows that the forward and backward traces of this measurement are identical, demonstrating that there is no hysteresis when the bottom gate is used as a fast axis (see also supplementary note 5). In these color-maps, traces of gate ineffectiveness appear in the central region where the CNP remains pinned, indicated by nearly vertical features.  The range in gate voltage of the top gate ineffectiveness evolves for different angular alignment, as shown in Figs. \ref{fig:Figure_2}a, c and e. Another characteristic of these measurements is the presence of jumps (or discontinuities) - marked by horizontal black arrows. This sudden change in the position of the CNP have been previously observed as a consequence of switching in the polarization of ferroelectric domains in parallel aligned BN systems \cite{yasudaStackingengineeredFerroelectricityBilayer2021}.

In contrast, sweeping the top-gate voltage as fast axis leads to hysteresis, as shown on Figs. \ref{fig:Figure_2}b, d and f, where we plot the subtraction (forward minus backward traces) of the four-probe resistance maps as a function of  $V_{\mathrm{tg}}/d_{\mathrm{tg}}$ (fast axis) and $V_{\mathrm{bg}}/d_{\mathrm{bg}}$ (slow axis). Each sweep direction of the map is colored accordingly.  A strong hysteretic behavior can be seen in Fig. \ref{fig:Figure_2}h when comparing the forward (red) and backward (blue) sweeping directions. Vertical regions of gate ineffectiveness are also observed in Fig. \ref{fig:Figure_2}b, d and f, where the position of the CNP do not change with the applied top gate voltage.

\begin{figure*}
    \centering
    \includegraphics[width = 1\textwidth]{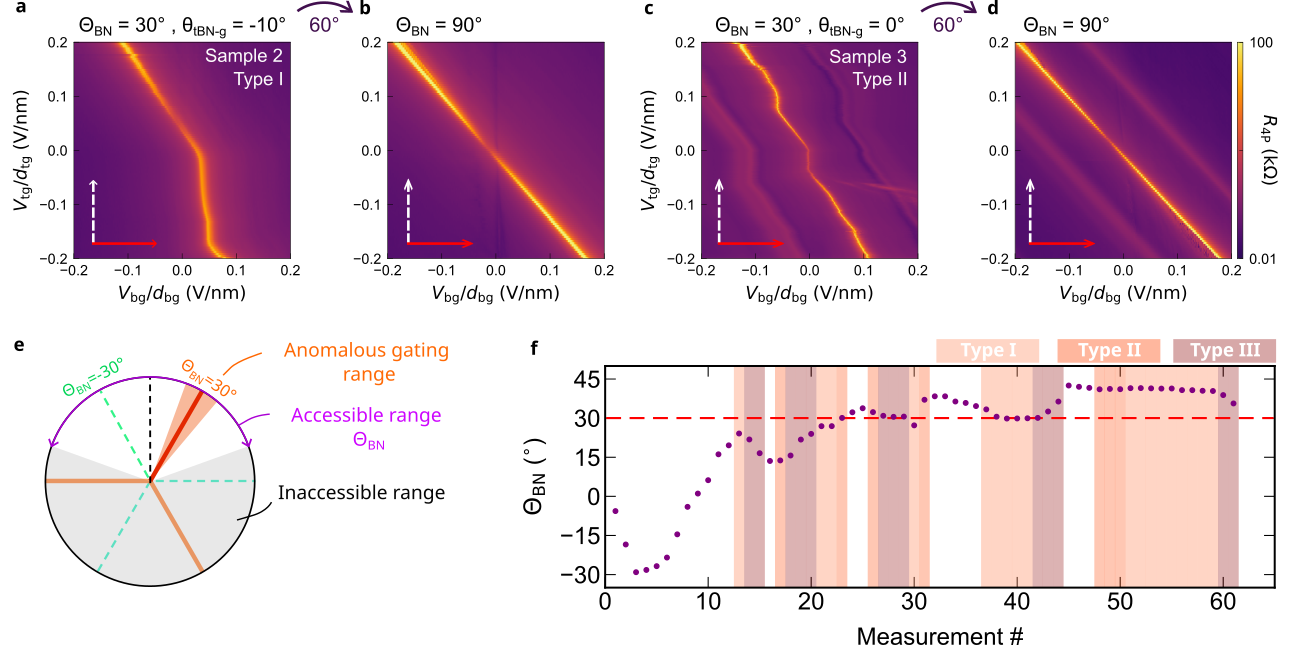}
    \caption{\textbf{Angular dependence of the anomalous gating.} \textbf{a}-\textbf{b}, Color-maps of the four-probe resistance as a function of the normalized bottom (fast axis) and top gate (slow axis) for sample 2 at $\Theta_{\mathrm{BN}}=30^{\circ}$ and $\Theta_{\mathrm{BN}}=90^{\circ}$, respectively. \textbf{c}-\textbf{d}, Color-maps of the four-probe resistance as a function of the normalized bottom (fast axis) and top gate (slow axis) for sample 3 at $\Theta_{\mathrm{BN}}=30^{\circ}$ and $\Theta_{\mathrm{BN}}=90^{\circ}$, respectively. Measurements taken at 6 K. All color-maps share the same color scale indicated on figure d. \textbf{e}, Representation of the angular alignments between the two BN layers where the gate ineffectiveness is observed. in the white region the dashed green and black lines represent $\Theta_{\mathrm{BN}}=-30^{\circ}$ and $0^{\circ}$, no anomalous gating is observed here. Red solid line and red shaded area represent the angular span where the anomalous gating is observed. Gray shaded area represents inaccessible range of angular alignment, red solid (green dashed) lines in this are represent angles at which we  do (not) expect to observe anomalous gating. \textbf{f}, Angular alignment as a function of the measurement number for sample 1. Color bars represent the different types of anomalous gating observed.} 
    \label{fig:Figure_4}
\end{figure*} 

The large hysteresis and gate ineffectiveness combine to form a closed loop, which also evolves with the angular alignment, as seen in Figs. \ref{fig:Figure_2}b, d and f.  These details hint that the phenomenon continuously evolves with the relative angle between the two BN layers. 

To summarize, type I anomalous gating has three particular features: i) strong gate asymmetry, ii) regions of top gate ineffectiveness that are observed when the top or the bottom gate are used as fast axis and iii) large hysteretic effect observed for the top gate as a fast axis. This type of anomalous gating has also been observed in bilayer graphene \cite{zheng_unconventional_2020,niu_giant_2022} encapsulated with BN, bilayer graphene encapsulated with BN and WSe2 \cite{niu_ferroelectricity_2025} and twisted graphite \cite{waters_anomalous_2025}.


\noindent \textbf{Anomalous gating type II:} this type shows both gate ineffectiveness and hysteresis but in a different way than type I. In Fig. \ref{fig:Figure_3}a, we show the color-maps of the four-probe resistance as a function of the normalized bottom (fast axis) and top (slow axis) gate for $\Theta_{\mathrm{BN}}=32^{\circ}$. In this configuration we do not observe gate ineffectiveness, however, a small amount of hysteresis is still observed, Fig. \ref{fig:Figure_3}b. When sweeping the top gate as fast axis, Fig. \ref{fig:Figure_3}c, we can see gate ineffectiveness on the edges of our gate voltage range. In other samples measured at low temperatures over an extended range of gate voltages, we observe top gate ineffectiveness when sweeping either the top or bottom gate as the fast axis, see supplementary note 6. Comparing these two samples suggests that the area of the parallelogram observed for type I is now too small to clearly resolve the hysteresis loop, at least at room temperature. It appears as though the loop is collapsing along its two oblique sides, perhaps as a continuation of the angular effects observed in Fig.\ref{fig:Figure_2}b, d and f.

Concerning the hysteretic behavior, we find it also here, as seen in the color-maps of Fig. \ref{fig:Figure_3}a and c, as well as in the line-cuts of Fig. \ref{fig:Figure_3}b and d. We find that the hysteresis is less pronounced  compared to type I (Fig. \ref{fig:Figure_3}h). However, for this type we can see it independently of the gate that is swept as fast axis. Another particularity of type II is that while both gates present hysteresis, the top gate presents the resistance peak of the CNP in an advanced fashion, Fig. \ref{fig:Figure_3}b, while the bottom gate presents it in a delayed way, Fig. \ref{fig:Figure_3}d. 


Type II anomalous gating has two particular features: i) top gate ineffectiveness regions observed when the top or the bottom gate are used as fast axis and ii) hysteretic effect observed for the bottom and top gate, with an advanced and delayed character, respectively. This type of anomalous gating has been previously reported in monolayer, bilayer and trilayer graphene\cite{zhang_electronic_2024}, twisted double bilayer \cite{chenAnomalousGatetunableCapacitance2024} and more complex graphene based structures \cite{wang_ferroelectricity_2022}. 

\noindent\textbf{Anomalous gating type III:}  The most striking feature of this type, measured for $\Theta_{\mathrm{BN}}=28^{\circ}$, is that while the bottom gate gives a standard gating behavior, without any hysteresis, the top gate shows a minimal effectiveness with a hysteretic behavior, Fig. \ref{fig:Figure_3}e-h.  We hypothesize that a larger electric field might be required to activate the effect of the top gate, extending much further than the gate voltage applied in our experiments. However, the same behavior was measured at low temperature where we have doubled the voltage range of our gates, with no change, see supplementary note 6.

This behavior looks like an extreme example of the behavior observed at $\Theta_{\mathrm{BN}}=41^\circ$ where the hysteresis loop, observed when the top gate is used as fast axis, is completely closed and both vertical sides of the parallelogram are on top of each other.

Type III anomalous gating has therefore three particular features: i) strong gate asymmetry, ii) almost full top gate ineffectiveness and iii) slight hysteretic effect observed for the top gate. We are not aware of any reports of this behavior in the literature.


Another striking outcome of our experiments is the angular dependence of the anomalous gating effect. It can be seen in Fig. \ref{fig:Figure_1}b and c, where the anomalous gating is not 60$^{\circ}$ periodic. This remarkable feature has been reproduced in two other samples, Fig. \ref{fig:Figure_4}a-d, for which we have measured both alignments at $T=6$ K. The low temperature measurements help us to rule out a possible difference in the Curie temperature between the two alignments that could explain the lack of 60$^{\circ}$ periodicity of the anomalous gating. Given the observation of the anomalous gating in all the measured samples we anticipate that this effect will have a 120$^{\circ}$ periodicity, see Fig. \ref{fig:Figure_4}e. 

In Fig. \ref{fig:Figure_4}a-d we can also see that the crystallographic orientation of the graphene with respect to the BN layers does not play a role in the observation of any of the types of anomalous gating reported here. Currently we have observed anomalous gating in three samples with alignments of 10$^{\circ}$ (Fig. \ref{fig:Figure_1}), 50$^{\circ}$ (Figs. \ref{fig:Figure_4}a-b) and 60$^{\circ}$ (Figs. \ref{fig:Figure_4}c-d) between the graphene and the top BN. In all the samples the crystallographic alignment necessary to observe anomalous gating was close to $\Theta_{\mathrm{BN}}\approx30^{\circ}\pm15^{\circ}$. We hypothesize that either the BN layers can feel the presence of the other layer through the graphene, considered somewhat unlikely in view of previous studies of graphene encapsulated with WS$_2$ \cite{niu_ferroelectricity_2025} or twisted graphite \cite{waters_anomalous_2025}; or this phenomena is created at the interface where the two BN layers touch each other. In our devices this will be outside the graphene Hall bar. In static-angle devices \cite{lin_room_2025,zhang_electronic_2024,zheng_unconventional_2020,niu_giant_2022,singh_stacking-induced_2025,kleinElectricalSwitchingBistable2023,chenAnomalousGatetunableCapacitance2024,waters_anomalous_2025,niu_ferroelectricity_2025}, overlapping regions can be observed, for example, beneath the top gate electrode, where the two BN layers overlap directly without graphene in between. 

In Fig. \ref{fig:Figure_4}f, we report the angular alignment between the BN layers for all our measurements (sample 1) and the type of anomalous gating observed. Each measurement represents a rotation with the AFM, note that some of these rotations are very small ($\approx0.05^{\circ}$), see supplementary note 3.  While trying to map out the boundaries of the anomalous gating range we remarked that the three types of behaviors do not appear in a particular sequence as a function of the angular orientation and can be found at different angles inside a $\pm15^{\circ}$ range around $\Theta_{\mathrm{BN}}\approx30^{\circ}$. Different types can also be found at very similar angles and particularly ``OFF zones" (white areas), where both gates have a standard behavior, can be found inside this range. This points to a possible role of translational movement, suggesting a sliding mechanism. This is also supported by the resistance jumps observed for type I and the fact that in all of our devices the gate ineffectiveness is always present in the top gate, which is the one that is freely able to move.

The observation of these three distinct types of anomalous gating in the same sample with only a small change of the angular alignment is for the moment insufficient to draw a definitive conclusion of the mechanisms creating the phenomena. However, we can use our experimental results to clarify some hypotheses put forward in the current literature:

The first one, the possibility to switch ON/OFF the hysteresis and gate ineffectiveness effects clearly shows that this is not a disorder (stack fault) induced effect, contrary to the proposal presented in \cite{niu_ferroelectricity_2025}. Second,  all the different behaviors observed across the samples seem to belong to the same phenomena and can be observed in the same sample at different angular alignments. This points towards a universality to the different observations in graphene-based systems \cite{lin_room_2025,zhang_electronic_2024,zheng_unconventional_2020,niu_giant_2022,singh_stacking-induced_2025,kleinElectricalSwitchingBistable2023,chenAnomalousGatetunableCapacitance2024,waters_anomalous_2025,niu_ferroelectricity_2025}. And finally, the angular alignment between the graphene and BN does not seem to play a role in the observation of the anomalous gating effect. 

Since this phenomenon is not associated with the formation of a moir\'e superlattice between the graphene and BN layers, we can rule out moir\'e-induced effects on the electronic band structure of BBG \cite{zheng_unconventional_2020} and/or charge localization within the moir\'e potential \cite{zheng_electronic_2023}. As a consequence, this phenomenon most likely does not rely on correlated electrons mechanisms, supporting the conclusions of \cite{niu_giant_2022}. We cannot completely rule out a co-sliding motion as suggested by \cite{lin_room_2025}. However, if this is the mechanism at play here, this will not be related to a full crystallographic alignment between the two BN layers but to a long range commensurate angle. At these angles the misaligned system still forms a crystal, albeit one with larger lattice vectors \cite{Bistritzer2010Jun,Chari2016Jul,Koren2016Sep,Inbar2023Feb}. Nonetheless, this cannot explain the lack of $60^{\circ}$ 
 periodicity in the anomalous gating effect.

The gate ineffectiveness observed here can be associated with charge transfer from either the graphene or the gate electrodes to the BN, as previously found in ferroelectric materials \cite{pesic_built-bias_2018,xue_unraveling_2021,huang_competing_2024}. However, the microscopic origin of the possible ferroelectric state in the BN layer and the particular angular dependence demonstrated here remains to be explained.

\section*{Conclusions}


To conclude we have reported on the evolution of the gate ineffectiveness and hysteresis effect in bilayer graphene/BN devices. We have proved that these phenomena are not related to the presence of a moir\'e superlattice but instead to the alignment between the two BN layers. Additionally, we have reported on the existence of three different types of behavior in this system. Although we currently cannot provide a clear explanation of this effect we expect that our experimental results inspire further theoretical and experimental developments.

\section*{Methods}

\noindent \textbf{Sample fabrication:} All samples were fabricated by dry transfer of 2D materials, following the technique explained in \cite{ribeiro2018twistable}. A stack of BN/Graphene was deposited on top of a graphite bottom gate using the flip stack technique. This BN/Graphene stack has been intentionally misaligned. Angular alignment of the BN handle and graphene structure is achieved using an AFM (FX40 from Park instruments) in contact mode to push the BN handle structure, following the technique explained in \cite{ribeiro2018twistable,arrighi2023non,Farrar2025Jan}.

\noindent \textbf{Electron transport measurements:} Room temperature transport measurements were performed at 100 nA using lock-in amplifiers and low frequency inside of our AFM. Low temperature measurements were performed at 6 K using lock-in amplifiers at low frequency.

\bibliography{Bib,references_Ferro2}

\section*{Acknowledgments}

The authors acknowledge discussions with Qiong Ma, Zhiren Zheng, Pablo Jarillo-Herrero, Jean-Christophe Charlier, Viet-Hung Nguyen, Ulf Gennser, Nicolas Leconte and Cory Dean.
 This work was done within the C2N micro nanotechnologies platforms and partly supported by the RENATECH network and the General Council of Essonne. This work was supported by: ERC starting grant N$^{\circ}$ 853282 - TWISTRONICS (R.R-P.), the DIM-SIRTEQ project TOPO2D, the DIM QuanTIP project Q-MAG and TBGSym and IQUPS. R.R.-P. acknowledge the Flag-Era JTC project TATTOOS (N$^{\circ}$ R.8010.19) and the Pathfinder project ``FLATS'' N$^{\circ}$ 101099139.  K.W. and T.T. acknowledge support from the JSPS KAKENHI (Grant Numbers 21H05233 and 23H02052) and World Premier International Research Center Initiative (WPI), MEXT, Japan.

 \section*{Author Contributions Statement}
 
R.R.-P., G.M., L.F. and D.M. designed the experiment. G.M. and L.S.F. fabricated the devices and performed the electron transport experiments and analyzed the data.  T.T. and K.W. grew the crystals of hexagonal boron nitride. M.K. developed the automatic recognition software for angle calculation. All authors participated to writing the paper. 

\section*{Competing Interest Statement}

The authors declare no competing interests.

\clearpage

\end{document}